\documentstyle[preprint,aps]{revtex}
\begin{document}

\draft
\setcounter{page}{0}

\title{ $^{6}$He{} + $\alpha${} clustering in $^{10}$Be }
\vspace*{10mm}
\author{N. SOI\'{C}$^{1}$, S. BLAGUS$^{1}$, M. BOGOVAC$^{1}$,
S. FAZINI\'{C}$^{1}$, M. LATTUADA$^{2}$,
M. MILIN$^{1}$,\\
\mbox{D\raisebox{0.30ex}
{\hspace*{-0.75em}-}\hspace*{ 0.42em}}. MILJANI\'{C}$^{1}$,
D. RENDI\'{C}$^{1}$, C. SPITALERI$^{2}$, T. TADI\'{C}$^{1}$
and M. ZADRO$^{1}$}

\address{
$^{1}$  Ru\mbox{d\raisebox{0.75ex}
{\hspace*{-0.32em}-}\hspace*{-0.02em}}er
 Bo\v{s}kovi\'{c} Institute, Zagreb, Croatia
\\
$^{2}$  INFN-Laboratorio Nazionale del Sud and
Universit\`a di Catania,
Catania,
Italy}

\date{\today}
\maketitle

\begin{quote}
\end{quote}

\vspace{15mm}

\begin{quote}

 {\bf Abstract.} -- In a kinematically complete measurement of the
  $^{7}$Li{}($^{7}$Li{},$\alpha${}$^{6}$He{})$^4$He
 reaction at $E_{i}$ = 8 MeV it was observed that the $^{10}$Be{}
  excited states at 9.6 and 10.2 MeV decay by $^{6}$He{} emission.
 The state at 10.2 MeV may be a member of a rotational band
 based on the 6.18 MeV 0$^+$ state.
\end{quote}

\vspace{15mm}

 {\it Introduction.} --
 In this paper we present the results of a search for the
 $\alpha${}+$^{6}$He{} decay of $^{10}$Be{}.
  In the most recent  compilation \cite{1} of the energy
 levels of light nuclei the only explicitly shown
 mode of decay of the $^{10}$Be{} states between
 the thresholds for neutron (6.81 MeV) and triton
 (17.25 MeV) decay is neutron emission, although the
 channels  $^6$He+$\alpha${}  , 2$\alpha${}+2n, $^8$Be+2n,
 and $^5$He+$^5$He are open from
 7.41, 8.39, 8.48, and 10.17 MeV,
 respectively. However, in the studies of $^{11}$Li decay
 \cite{2} the delayed emission of $^{6}$He{}  was observed and
 it was attributed to the
 decay of the excited states of $^{10}$Be{}.
 Namely, it was claimed that the
 coincident $^{6}$He{}-$\alpha${} spectra could be explained by
 the $\beta^-$  decay of $^{11}$Li into the 10.6 and
 18.5 MeV  states of $^{11}$Be, which
 decay by neutron emission to the 9.4 and 11.6 MeV states
 in $^{10}$Be{} and  each of them
 then disintegrate into an $\alpha$-particle and a $^{6}$He{}.
 However,
 this is not the only way how the $^{6}$He{}+$\alpha${}+n
 final state
 can be reached. The $^{11}$Be high excited states can also
 decay into the $^{6}$He{}+$^5$He and $\alpha${}+$^7$He channels
 with subsequent disintegration of
 neutron unstable $^5$He and $^7$He nuclei.
 Taking into account experimental
 conditions in the $^{11}$Li decay measurements, the
 involvment of the $^{10}$Be{} states and their
 $\alpha${}+$^{6}$He{} decay
 cannot be unambigously claimed. Another indication of
 possible $^{6}$He{}+$\alpha${} decay of the
 $^{10}$Be{}  states came from the studies of the $^{7}$Li{}
 ($^{7}$Li{} ,$^{6}$He{})$^8$Be reaction \cite{3}.
 The $^{6}$He{} spectra
 could not be explained  exclusively by contributions from the
 sequential processes through
 different $^8$Be states. A broad structure
 in the total $^9$Be(n,$\alpha${})$^6$He reaction cross
 section \cite{4}, centered around 9.6 MeV in $^{10}$Be{},  may
 be another indication of the $\alpha${}+$^{6}$He{}
 clustering of the  states in this region.

 Nuclei in the middle of the 1p shell exhibit collective nature.
 Although the independent particle model can account for many
 features of the nuclei, there are many exceptions, like
 enhanced electromagnetic transitions, large quadrupole
 moments, "unexpected" low lying nonnormal
 parity states (like the $1/2^+$ ground state in $^{11}$Be),
 large rms radii and $\alpha$-decay widths etc.
 Some of these properties could be easily
 explained by the cluster structure of the nuclei. There
 have been  several theoretical studies of the structure
 of the A = 10 nuclei.
 Gabr and Hackenbroich \cite{5} chose the cluster functions
 of $^{10}$B  to belong to spatial symmetry [442], which
 correspond to $^8$Be core [44]
 and an extra deuteron or a $^6$Li core [42] and an extra
 $\alpha$-cluster. The intercluster relative motions were
 represented by a small number of
 Gaussian functions. Only positive parity states with
 low excitations were
 determined. In a two-$\alpha$-particle-plus-dinucleon cluster model
 by Nishioka \cite{6}, both $^{10}$Be{} and $^{10}$B states were
 calculated. The (1$_3^+$,0), (0$_2^+$,1), and (2$_3^+$,1)
 $^{10}$B level energies were  reproduced, which had not been
 the case  by any other model investigation.
 These states were found to have a well developed
 $^6$Li$_{g.s.}$+$\alpha${} or $^6$Li(0$^+$,1)+$\alpha${}
 cluster structure, respectively.  In another
 investigation the
 10-nucleon system was studied with the multiconfiguration
 and  multichannel  resonating group method \cite{7}.
 The model space employed  was spanned by the
 $\alpha${}+$^6$Li, d+$^8$Be, d+$^8$Be$^*$, and $\alpha${}+$^6$Li$^*$
 cluster configurations with $^6$Li$^*$ and $^8$Be$^*$
 being the rotational excited states of $^6$Li and $^8$Be having
  d+$\alpha${} and $\alpha${}+$\alpha${}
 cluster structure and L = 2. Bound and resonant levels
 obtained in such a way  correspond fairly well to
 the known low lying states.
 However, in all these theoretical studies the states at higher
 excitations ($>$ 9 MeV)  were not investigated.

 These states can be easily reached by the
 $^{7}$Li{}+$^{7}$Li{} reactions.
 The $^{7}$Li{}($^{7}$Li{},$\alpha${}$^{6}$He{})$^4$He
 reaction was chosen
 for the search of  the $\alpha${}+$^{6}$He{}
 cluster states in $^{10}$Be{}
 for the following reasons:
 i) it has high positive Q-value (7.37 MeV)
 allowing measurements at low energies;
 ii) only the well known $^8$Be states (0, 3.0,
 and 11.4 MeV) together with those from $^{10}$Be{}
 can contribute to the coincident spectra; iii) the
 complex nature of the reaction at low energies
 should be more suitable for the excitation of these special
 states than some "simple"
 reactions like (d,p), ($\alpha${},$^3$He) etc.

\vspace{10mm}
 {\it Experiment} --
  A 3 particle nA 8 MeV $^{7}$Li{} beam from the
 Ru\mbox{d\raisebox{0.75ex}
 {\hspace*{-0.32em}-}\hspace*{-0.02em}}er
 Bo\v{s}kovi\'{c} Institute EN Tandem Van de Graaff
 accelerator was used to bombard
 isotopically enriched $^{7}$Li{}F targets (100 - 320
 $\mu$g/cm$^2$)  evaporated on a thin carbon backing.
 Reaction products were observed
 with two solid state detector telescopes, each
 consisting of a thin $\Delta$E detector (9 $\mu$m) and
 a thick (280 $\mu$m) rectangular position sensitive
 detector (PSD).
 PSD covered an angle of 12$^o$ in horizontal and 1.5$^o$
 in vertical axis. Their horizontal angular resolution
  was better than 0.3$^o$.
 The telescopes were positioned  on the
 opposite sides with respect to the beam.
 The measurements  were performed for several
 setting angles between 40$^o$ and 65$^o$.
 The particle energy, "position" and
 energy loss pulses
 were recorded by a data acquisition system \cite{8}.
 From these measured values the energy-momentum
 (EP) plots as well as Q-value spectra were made
 \cite{8a}. Other details on the experiment and
 analysis can be found in \cite{9}.

\vspace{10mm}
 {\it Results and discussion} --
  Fig. \ref{fig1} shows measured Q-value spectra
 for the $^{6}$He{}-$\alpha${}
 and $\alpha${}-$\alpha${}  coincidences. In the case of
 $^{6}$He{}-$\alpha${}
  coincidences background is very small,
 but in the second case
 the contributions  from the
 $^{19}$F($^{7}$Li{},$\alpha${}$\alpha${})$^{18}$O
 and $^{12}$C($^{7}$Li{},$\alpha${}$\alpha${})$^{11}$B
 reactions are also present.
 In this case, in addition to the $^{6}$He{} ground state peak,
 five-body (3$\alpha$+2n) continuum starting at
 Q = 6.4 MeV together with a peak
 corresponding to the $^6$He first excited state (E$_x$ = 1.8 MeV)
 are also visible. Other structures cannot be unambigously
 attributed to the $^{6}$He{} states.

 The $\alpha${}-$^{6}$He{}-$\alpha${} events have also been
 sorted into three relative energy plots. The
 $^{6}$He{}-$\alpha${} coincidences, measured at setting angles
 $\Theta_1$=$\Theta_2$=45$^o$ ($\Delta \Phi$ = 180$^o$),
 are displayed in the first three plots of Fig. \ref{fig2}. Two
 prominent groupings are observed: one corresponding to the
 excitation energies in the $^{6}$He{}-$\alpha${} system
 (E$_{13}$) between 2.5 and 3 MeV and the other corresponding
 to the energies in the $\alpha${}-$\alpha${} system (E$_{23}$)
 around 3 MeV i.e.  to the first excited state of $^{8}$Be.
 Although a wider range of the $^{10}$Be excitation energies
 was covered in the experiment, we concentrate here only on
 the part of the spectra below 4 MeV in the
 $\alpha${}-$^{6}$He{} relative energy, which is not
 affected by the sequential processes in other two pairs.

 From the recorded data the $\alpha${}-X coincidences
 (X being heavy particles with A $\geq$ 7)
 are also selected with the aim
 to obtain information on the $\alpha${}-$^{9}$Be-n
 contributions. These heavy particles, detected in the
 $\Delta$E-detectors, were not identified - only their
 energy was measured. With an additional requirement
 that the  events fall into the allowed kinematical
 regions for the
 $^{7}$Li{}($^{7}$Li{},$\alpha${}$^{9}$Be)n reaction,
 they are sorted into the E$_{ij}$-$\Theta_{\alpha}^{L}$
 plots.  E$_{ij}$ are the
 relative energies in the $^{9}$Be-n pairs, calculated
 directly from the energy and detection angle
 of $\alpha${}-particles, $\Theta_{\alpha}^{L}$.
 An example is shown on the fourth plot of
 Fig. \ref{fig2}.
 The vertical structures seen in the plot obviously
 correspond to the sequential processes through the
 neutron decaying states of $^{10}$Be{}. The
 ``background'' is due to the sequential processes
 through the $^{5}$He and $^{13}$C states from the
 same  $^{7}$Li{}($^{7}$Li{},$\alpha${}$^{9}$Be)n
 reaction as well as to the reactions on $^{19}$F.
 (The low energy cut was made in order to avoid
 random coincidences caused by the $^{7}$Li
 elastic scattering).

   Fig. \ref{fig3} shows the results of the experiment:
 the $^{10}$Be{} excitation energy
 spectra   from the
 $^{7}$Li{}($^{7}$Li{},$\alpha${}$^{6}$He{})$^4$He
 and $^{7}$Li{}($^{7}$Li{},$\alpha${}$^9$Be)n
 reactions.
 The uncertainties of peak positions and the resolution in
 excitation energy were estimated by the peaks corresponding
 to the ($^7$Li,$\alpha$) reactions to bound states of
 $^{10}$Be. Their values were found to be $<$100 keV and
 250 keV, respectively.

 In the $\alpha${}-$^9$Be spectrum
 two  structures are visible. The first one corresponds to a
 doublet of $^{10}$Be{} states (9.27 and 9.64 MeV) and the second
 one to a state at 10.57 MeV, all
 of them previously observed in different processes.
 Until recently the value of 9.4 MeV was quoted
 for the energy of the second member of the
 doublet. The present
 measurements support recent findings from the
 study of the $^{7}$Li{}($\alpha${},p)$^{10}$Be{}  reaction \cite{10}
 that the energy is somewhat higher, i. e. 9.64 MeV.
 This state also decays into $\alpha${}+$^{6}$He{} channel,
 which supports previous claims about the
 involvement of this state in one of the final stages
 of the $^{11}$Li decay.

 The other spectrum, $^{6}$He{}-$\alpha${} and $\alpha${}-$\alpha${}
 coincidencies,  has a
 distinctive  peak centered at 10.2 MeV.
  The width of this state is less than 400 keV.
 There hasn't been any mention
 of a state at this energy except for
 the $^{7}$Li{}($\alpha${},p)$^{10}$Be{} measurement \cite{10}.
 A double peaked structure in this energy region can also be
 seen in an $\alpha${}-particle  spectrum from the
 $^{7}$Li{}($^{7}$Li{},$\alpha${})$^{10}$Be{} reaction measured at
 30.3 MeV \cite{11}. This state does not decay into n+$^9$Be
 channel, which explains very well why it was not
 observed in any
 neutron transfer reaction on $^9$Be \cite{12}.
  One can also mention that its energy coincides with the
 threshold for $^{10}$Be{} disintegration into two $^5$He$_{g.s.}$.
  It is interesting to note here that the proton
 angular distributions
 from the $^{7}$Li{}($\alpha${},p)$^{10}$Be{} reaction for the 10.2 and
 11.8 MeV states are almost identical in shape \cite{10}.
 If the 11.8 MeV state is the
 4$^+$ member of the ground state rotational band, as claimed,
 then from the similarity of the angular distributions it
 follows that the 10.2 MeV state should at least have the
  positive parity, and maybe  even the spin of 4.
 One can then further speculate that this state may
 be the 4$^+$ member of the rotational band based on
 the 6.18 MeV excited 0$^+$ state with its
  2$^+$ state at 7.54 MeV.
 Its excitation (10.2 MeV) is close to the energy (10.7 MeV)
 expected from the J(J+1) rule and the energy
 difference (1.36 MeV) between the 0$^+$ and 2$^+$ states.
 These two states, both in $^{10}$Be{} and $^{10}$B, are known to
 have well developed $^{6}$He{}+$\alpha${}  and $^6$Li(0$^+$,1)+$\alpha${}
 cluster structure, respectively (see e. g. \cite{6}).
 Following the sequence of these states in $^{10}$Be{}
 (6.18, 7.54, 10.2 MeV) and
 the first two in $^{10}$B (7.56, 8.89 MeV)
  one may expect that the $^{10}$B  level at 11.5 MeV, the only
 well established state between 10.9 and 12.5 MeV,
 is the third member of this band.
  Small energy separation between these states
 would then  imply a large moment of inertia, i. e.
 they would be very extended nuclear systems.
 Because the state at 10.2 MeV decays by emission
 of $^{6}$He{}, the well established two-neutron halo
 nucleus (see e. g. \cite{13}), and because other
 two states (6.18 and 7.54 MeV) have also the
 $\alpha${}+$^{6}$He{} structure, one may ask what is the
 relation between these states and the two-neutron halo
 states, which may be expected in $^{10}$Be{} close to
 the 2n emission threshold (8.48 MeV).

 To conclude: the existence of the poorly known $^{10}$Be{} state
 at 10.2 MeV is confirmed. This state decays into the
 $\alpha${}+$^{6}$He{} channel, but not into n+$^9$Be.
 Together with other two states (0$^+$ at 6.18 MeV and
 2$^+$ at 7.54 MeV) it may make a rotational band
 of a very extended nuclear system. Further studies of
 the states   in all three A = 10 nuclei  having T=1 states,
 $^{10}$Be{}, $^{10}$B, and $^{10}$C,
 would give clear answer about the possible
 existence of the peculiar nuclear systems.

\begin{center}
* * *
\end{center}

 The authors wish to thank Dr. M. Jak\v{s}i\'{c} and the staff
 of the Tandem accelerator for their help during the
 experiment.

\begin{figure}
\caption{-- Q-value spectra for the ($^7$Li,$\alpha$$^6$He)
 and ($^7$Li,$\alpha$$\alpha$) reactions on a $^7$LiF
 target and its carbon backing.}
\label{fig1}
\end{figure}

\begin{figure}
\caption{-- The E$_{ij}$-E$_{jk}$ plots of the
 $^6$He-$\alpha$ coincidencies and the
 E$_{ij}$-$\Theta_{\alpha}^{L}$ plot of the
 X(heavy particle)-$\alpha$ coincidencies, all measured
 at E$_{i}$ = 8 MeV and $\Theta_{1}$ =
 $\Theta_{2}$ = 45$^o$ ($\Delta \Phi$ = 180$^{o}$).}
\label{fig2}
\end{figure}

\begin{figure}
\caption{-- $^{10}$Be excitation spectra obtained from the
$^7$Li($^7$Li,$\alpha$$^6$He)$^4$He and
$^7$Li($^7$Li,$\alpha$$^9$Be)n reactions at  E$_i$ = 8 MeV
(large difference in the number of events for these two
 processes is mainly due to the larger $^9$Be detector
 solid angle).}
\label{fig3}
\end{figure}

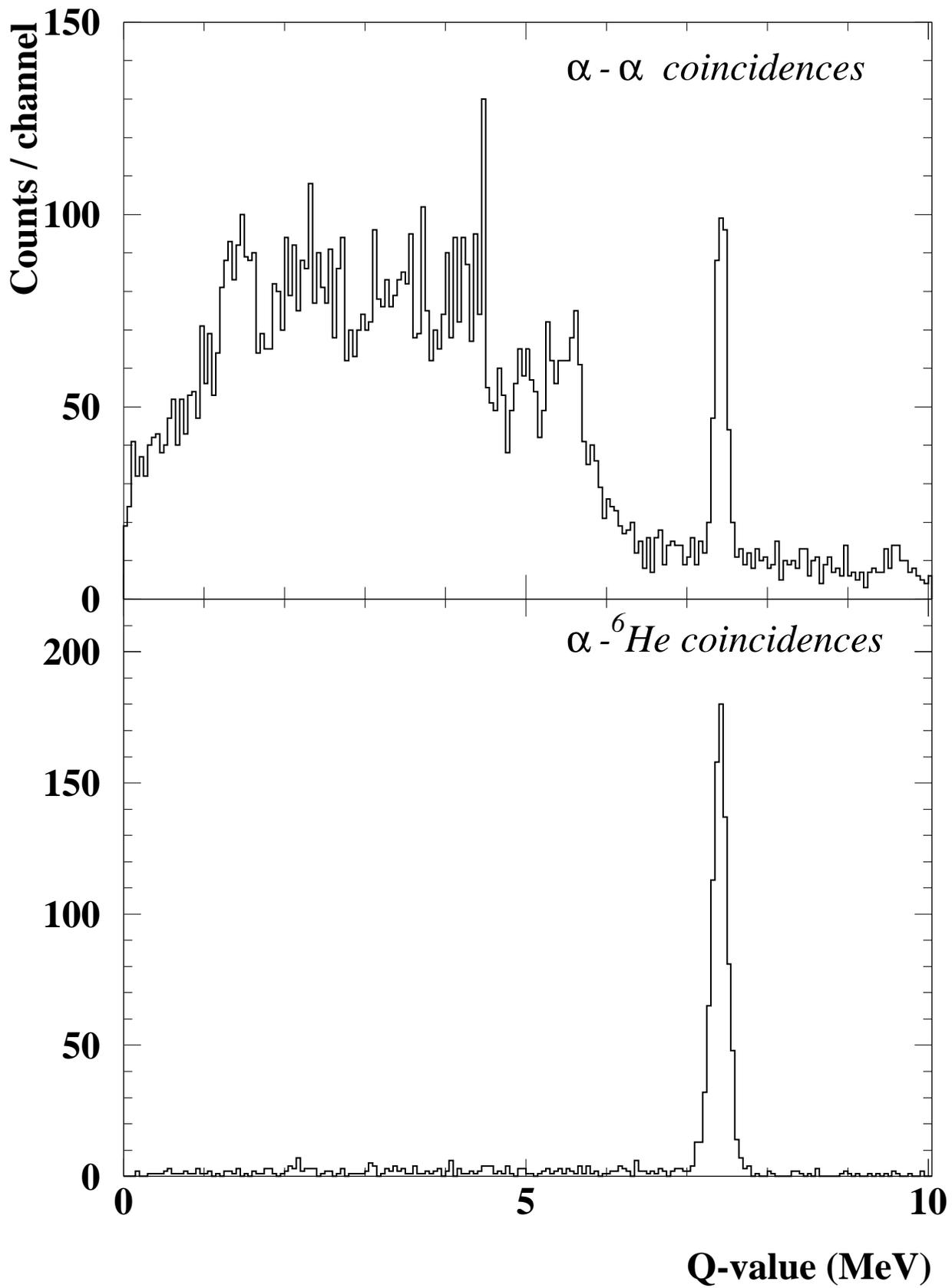
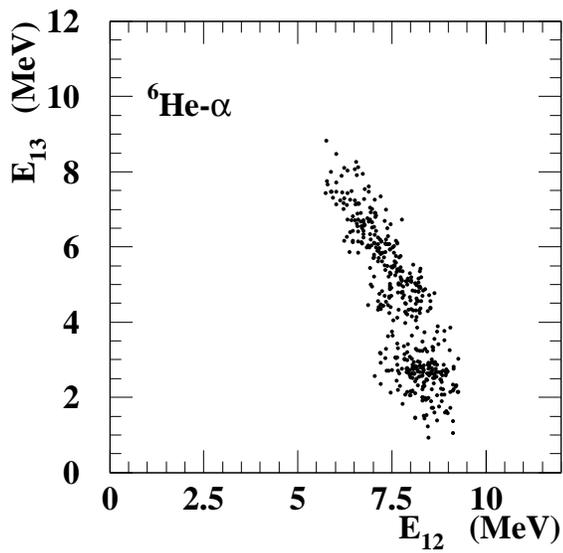
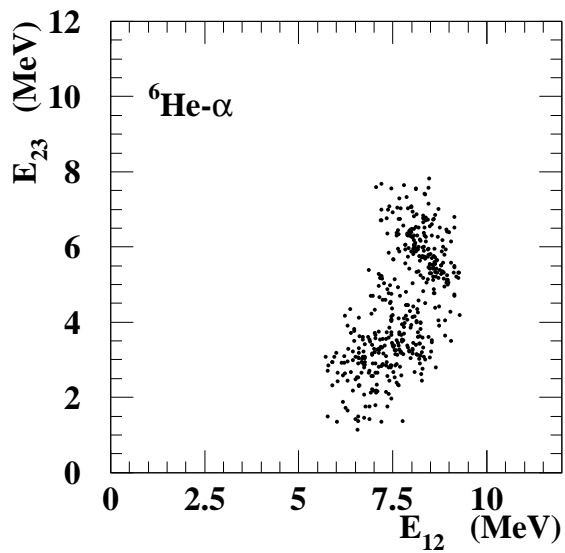
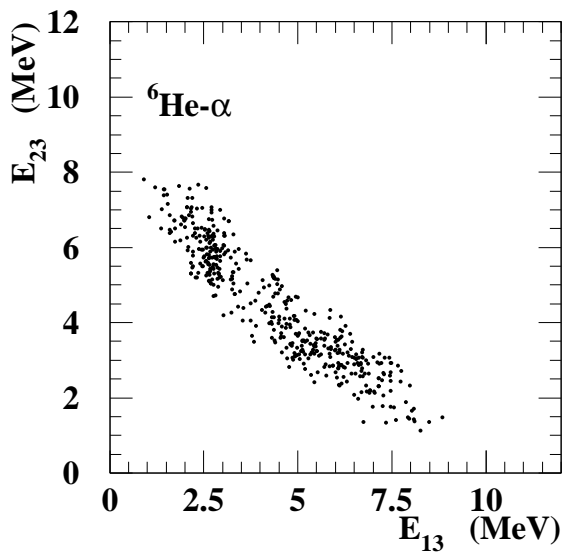
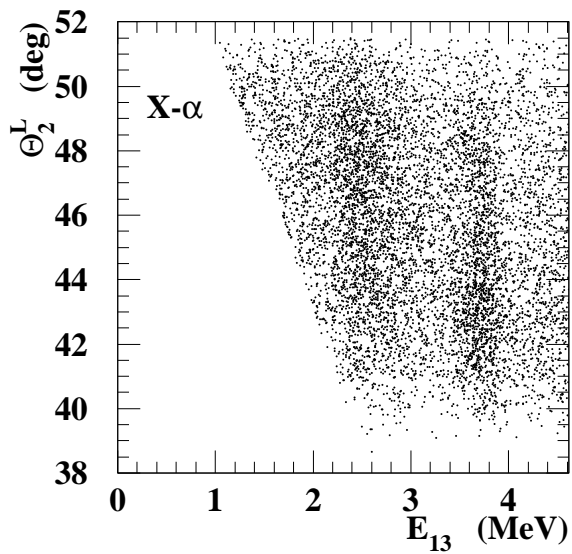
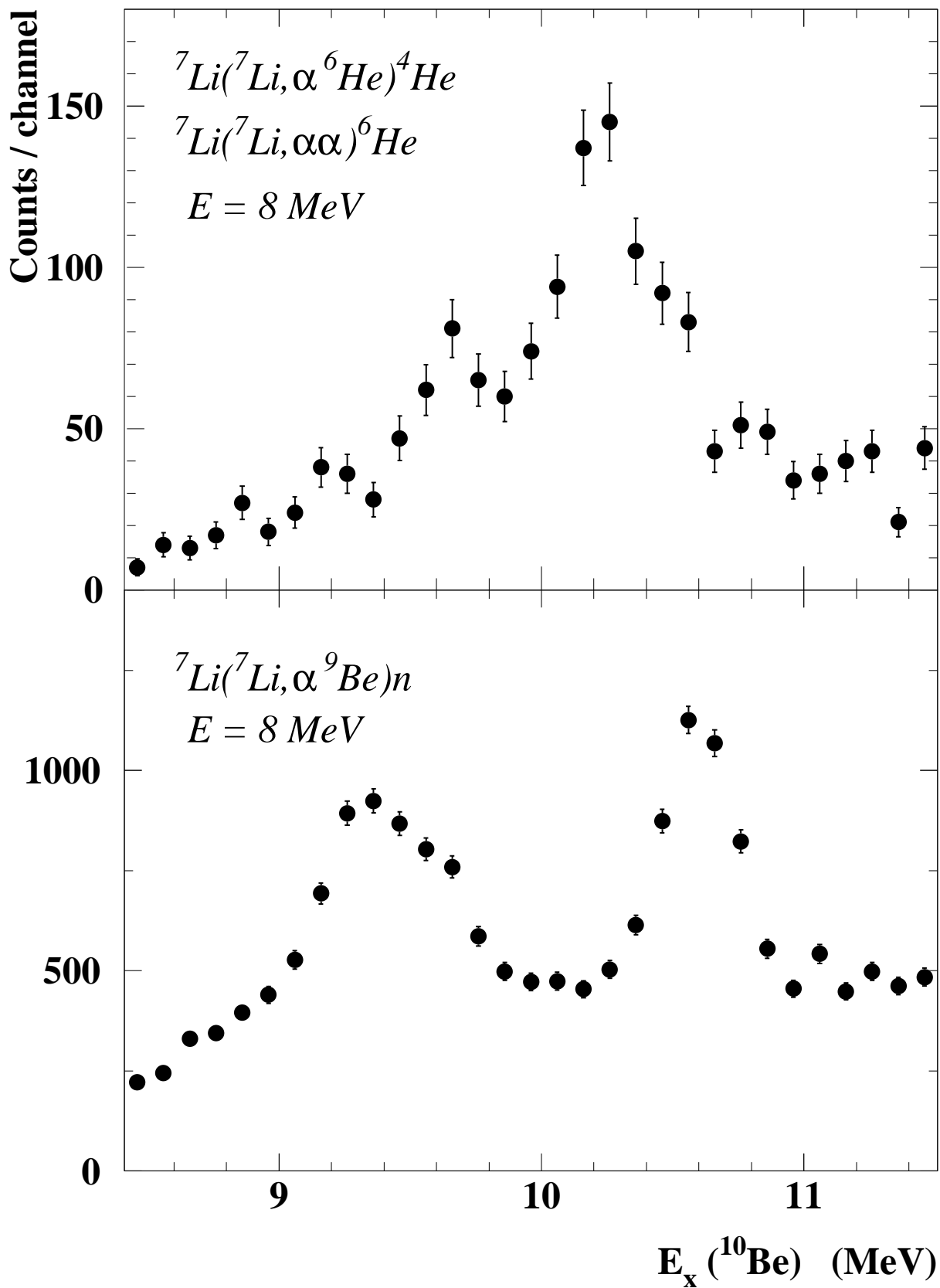
\end{document}